\documentclass[runningheads]{llncs}
\usepackage[T1]{fontenc}
\usepackage{graphicx}
\usepackage{cite}
\usepackage{marvosym}
\usepackage{ifsym}
\usepackage{multirow}
\usepackage{bm}
\usepackage{amsmath}
\usepackage{hyperref}
\hypersetup{hypertex=true,
	colorlinks=true,
	linkcolor=blue,
	anchorcolor=blue,
	citecolor=blue}

\begin{document}
%

\title{Fusing Structural and Functional Connectivities using Disentangled VAE for Detecting MCI}

%
%
\author{Qiankun Zuo\inst{1,8} \and
Yanfei Zhu\inst{2}(\textrm{\Letter}) \and
Libin Lu\inst{3} \and
Zhi Yang\inst{4} \and
Yuhui Li\inst{5} \and
Ning Zhang\inst{6,7}}

%
\authorrunning{Zuo et al.}
%

\institute{School of Information Engineering, Hubei University of Economics, Wuhan 430205, China\\
	\and School of Foreign Languages, Sun Yat-sen University, Guangzhou 510275, China\\
	\and School of Mathematics and Computer Science, Wuhan Polytechnic University, Wuhan 430023, China\\
	\and College of Electronics and Information Engineering, Sichuan University, Chengdu 610065, China\\
	\and Goertek Inc., Beijing 100083, China\\
	\and Beijing SmartDoor Technology Co., Ltd., Beijing 101399, China\\
	\and Beijing Zhongke Ruijian Technology Co., Ltd., Beijing 100088, China\\
	\and Zhuhaishi Jiexinsoftware Technology Co., Ltd., Zhuhai 519090, China\\
	\email{Corresponding email: zhuyf53@mail.sysu.edu.cn}
}

\maketitle              
\begin{abstract}
Brain network analysis is a useful approach to studying human brain disorders because it can distinguish patients from healthy people by detecting abnormal connections. Due to the complementary information from multiple modal neuroimages, multimodal fusion technology has a lot of potential for improving prediction performance. However, effective fusion of multimodal medical images to achieve complementarity is still a challenging problem. In this paper, a novel hierarchical structural-functional connectivity fusing (HSCF) model is proposed to construct brain structural-functional connectivity matrices and predict abnormal brain connections based on functional magnetic resonance imaging (fMRI) and diffusion tensor imaging (DTI). Specifically, the prior knowledge is incorporated into the separators for disentangling each modality of information by the graph convolutional networks (GCN). And a disentangled cosine distance loss is devised to ensure the disentanglement's effectiveness. Moreover, the hierarchical representation fusion module is designed to effectively maximize the combination of relevant and effective features between modalities, which makes the generated structural-functional connectivity more robust and discriminative in the cognitive disease analysis. Results from a wide range of tests performed on the public Alzheimer's Disease Neuroimaging Initiative (ADNI) database show that the proposed model performs better than competing approaches in terms of classification evaluation. In general, the proposed HSCF model is a promising model for generating brain structural-functional connectivities and identifying abnormal brain connections as cognitive disease progresses.

\end{abstract}

\section{Introduction}

Alzheimer's disease (AD) is one of the most prevalent progressive and irreversible degenerative disorders affecting the elderly, where the initial onset is considered to be mild cognitive impairment (MCI). Memory loss, aphasia, and other declining brain functions represent MCI-related symptoms and are indicative of pathological changes \cite{tsentidou2019cognition}. According to literature \cite{davatzikos2011prediction}, the AD conversion rate from MCI is much higher than in normal people. In addition to making AD patients more depressed and anxious, it also lowers their quality of life and places a heavy financial burden on their families due to the high cost of care \cite{peres2019oral}. Furthermore, there is still no effective treatment for the illness \cite{keren2017unique}. Early diagnosis and treatment of patients with MCI can effectively slow their progression to AD. Therefore, developing an effective machine learning model for analyzing scanned medical imaging and other field applications for disease detection has attracted growing attention \cite{ml3,shen2016subcarrier,gibson2018niftynet,ml2,wang2018skeletal,hong2020brain,ml4,shgan4,shgan2,zeng2017ga,zuo2023diffusion}.

When the human brain completes a certain task, multiple brain regions need to interact with each other, so studying cognitive diseases from the perspective of brain connectivity is more explanatory. Brain networks are based on graph theory, where nodes usually represent neurons or regions of interest (ROIs), and edges represent the relationships between nodes (i.e., brain regions)\cite{zuo2021multimodal}. Disease-related information can be conveyed in various ways by multiple modal images\cite{sh2,sh3,hong2022unsupervised,lei2022predicting}. fMRI (functional magnetic resonance imaging) records brain activity and can reveal abnormal functional connectivity (FC) associated with disease \cite{hirjak2020multimodal,FC2}. White matter fiber bundles in the brain can be recorded using diffusion tensor imaging (DTI), which can reveal abnormal structural connectivity (SC) between different brain regions \cite{Honey,SC2}. Compared with the traditional imaging-based method in MCI diagnosis\cite{mci1,ml5,mci2,shgan5,shgan6,ml6}, the connectivity-based methods show superior performance in accuracy evaluation by graph convolutional networks (GCN)\cite{lei2021diagnosis,zuo2023diagnosis}. Researchers either use SC or FC to perform an early AD diagnosis clinically. For example, Zuo et al.\cite{zuo2022constructing} designed a transformer-based network to construct FC from functional MRI and improve MCI diagnosis accuracy compared with empirical methods. Since both fMRI and DTI can explore complementary information in patients, multimodal fusion has produced superior results in MCI diagnosis\cite{zuo2021prior,zong2022multiscale,hong2022source}. The work in \cite{mci3} has proved the success of fusing SC and FC in MCI prediction. They utilized the local weighted clustering coefficients to adaptively fuse the functional and structural information, thus enhancing the disease diagnosis. This shows that fusing multimodal brain networks is promising and is becoming a hot topic in cognitive disease analysis\cite{shgan3,cls1,zhang2021deep}. However, the information from one modality may act as noise to prevent the expression of the other modality in previous approaches, which always combine the disentangled information of multimodal information. Consequently, minimizing the components that can have a detrimental impact on one another during the fusion process is the key to efficiently merging DTI and fMRI data.

The variational autoencoder (VAE) is one of the most generative methods \cite{bftd1,mo2009variational,kingma2019introduction} in information fusion by encoding features into latent representations,and the graph convolutional network (GCN) has a strong advantage in constructing topological features. Inspired by the observations, in this paper, a novel hierarchical structural-functional connectivity fusion (HSCF) model is proposed to construct brain structural-functional connectivity matrices and predict abnormal brain connections based on functional magnetic resonance imaging (fMRI) and diffusion tensor imaging (DTI). The main advantages of this paper are the following: (1) The prior knowledge is incorporated into the separators for disentangling each modal information by GCN, which can separate the connectivity information in topological space and is more suitable for downstream fusion. (2) The hierarchical representation fusion module is designed to effectively maximize the combination of relevant and effective features between modalities, which makes the generated structural-functional connectivity more robust and discriminative in the cognitive disease analysis. Comprehensive results on the Alzheimer's Disease Neuroimaging Initiative (ADNI) database show that the performance of the proposed model outperforms other competitive methods in terms of classification tasks.

\section{Proposed Method}
\subsection{Disentangled VAE}

The input to our framework is the graph data, where nodes represent the ROIs and edges represent the SC or FC. To simplify the description, we denote the SC and FC as the $\bm{A_1}$ and $\bm{A_2}$ respectively. Both SC and FC have the dimension size $N \times N$. The $N$ represents the total number of brain regions studied in our study. The prior knowledge refers to the relative volume of anatomical brain regions, and we construct the node feature (NF) by translating each ROI's volume into a one-hot vector. The NF is denoted as $\bm{X}$ with a size of $N \times N$.

\begin{figure}[htbp]
	\includegraphics[width=\textwidth]{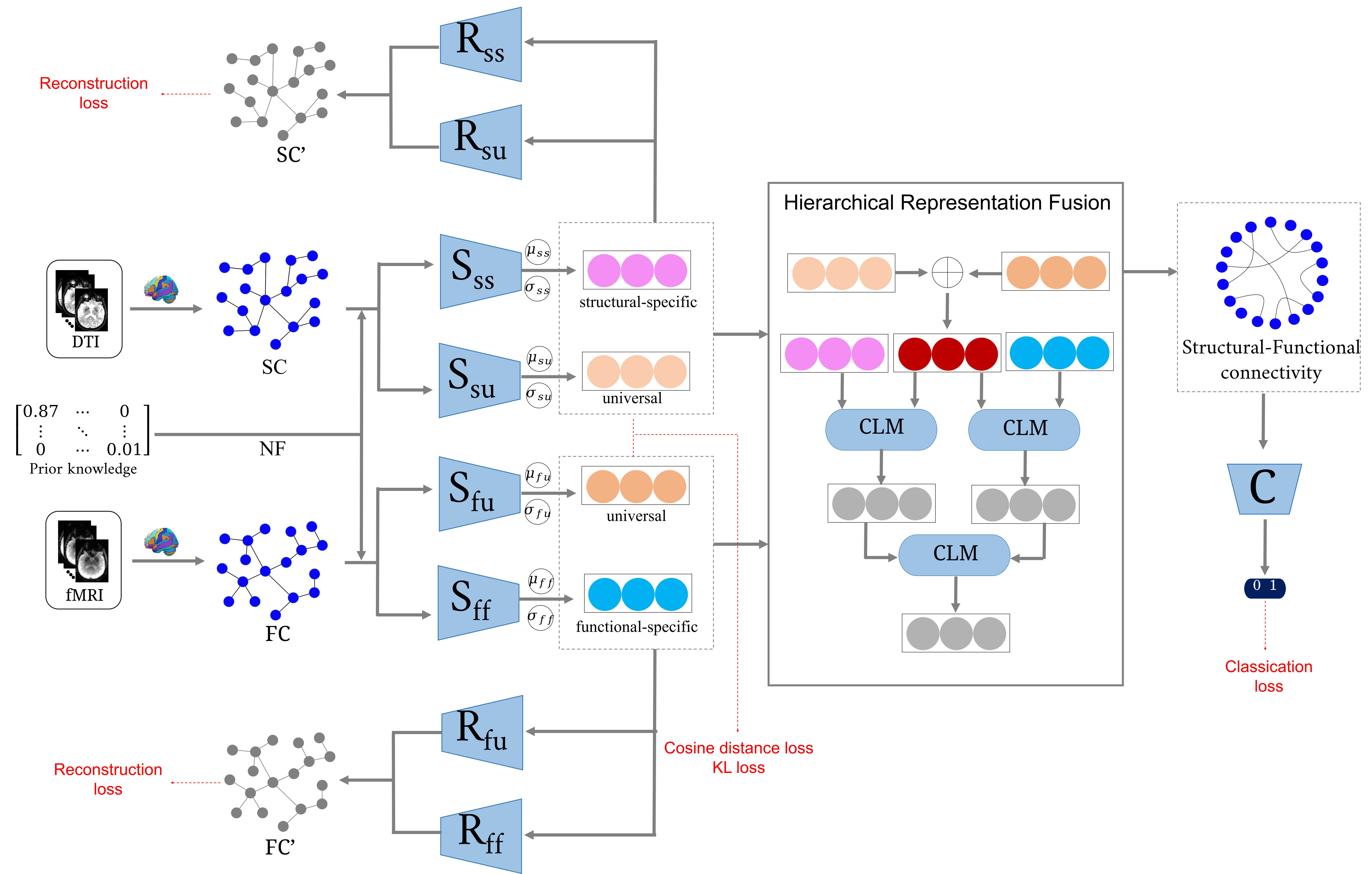}
	\caption{The framework of the proposed HSCF using DTI and fMRI. It consists of three parts: the encoders, the decoders, and the hierarchical representation fusion.
	} \label{fig1}
\end{figure}

The framework is shown in Fig.~\ref{fig1}. The disentangled VAE consists of four separators and four reconstructors. As an example, consider the distangled structural connectivity. The two separators are $S_{ss}$ and $S_{su}$, where each of them takes $\bm{A_1}$ and $\bm{X}$ as input and outputs the latent variables. The difference is that the former learns the structural-specific component ($\mu_{ss}, \sigma_{ss}$), while the latter learns the universal component ($\mu_{su}, \sigma_{su}$). The network structure of them contains three GCN layers: the first two layers have hidden dimensions of 64 and 32, respectively; the last layer has hidden dimensions of 16. Except for the last layer, the $ReLU$ activation function is applied to all GCN layers. The computation procedure can be defined as follows:
\begin{equation}
\mu_{ss}, \sigma_{ss} = S_{ss} (\bm{A_1}, \bm{X}); \mu_{su}, \sigma_{su} = S_{su} (\bm{A_1}, \bm{X})
\end{equation}
\begin{equation}
\mu_{ff}, \sigma_{ff} = S_{ff} (\bm{A_2}, \bm{X});  \mu_{fu}, \sigma_{fu} = S_{su} (\bm{A_2}, \bm{X})
\end{equation}
here, each pair of latent variables has the same dimension $N \times 16$. The latent variable pairs can be considered a standard normal distribution, where we can obtain the latent representations by sampling operations. Supposing that the latent representations are $\bm{Z_{ss}}$, $\bm{Z_{su}}$, $\bm{Z_{ff}}$, and $\bm{Z_{fu}}$, we can recover the SC and FC by the reconstructors. The structure of the reconstructor has an N filter with a kernel size of $16 \times 1$. The final output is the matrix inner product, followed by a $sigmoid$ activation function. The fomulua can be expressed by:

\begin{equation}
\bm{A_1'} = 0.5(\bm{A_{s1}}+\bm{A_{s2}})
\end{equation}
\begin{equation}
\bm{A_{s1}} = R_{ss} (\bm{Z_{ss}}), \bm{A_{s2}} = R_{su} (\bm{Z_{su}})
\end{equation}
\begin{equation}
\bm{A_2'} = 0.5(\bm{A_{f1}}+\bm{A_{f2}})
\end{equation}
\begin{equation}
\bm{A_{f1}} = R_{ff} (\bm{Z_{ff}}), \bm{A_{f2}} = R_{fu} (\bm{Z_{fu}})
\end{equation}

\subsection{Hierarchical Representation Fusion}
The disentangled representations are combined to generate structural-functional connectivity for fusing complementary information. The hierarchical representation fusion (HRF) consists of three stages: (1) fusing the universal representations to obtain phase-1 representation; (2) partially fusing the phase-1 representation with modality-specific representations using connectivity linear mapping (CLM); this stage outputs phase-2 representations; and (3) continuing to incorporate the phase-2 representations to obtain phase-3 representation. The CLM consists of a two-layer multilayer perceptron (MLP). The output dimension of each layer is the same as the latent variable. The generated structural-functional connectivity is defined as:

\begin{equation}
\bm{A_m} = SFC = HRF(\bm{Z_{ss}},\bm{Z_{su}},\bm{Z_{ff}},\bm{Z_{fu}})
\end{equation}
The classifier $C$ shares the same structure with the work in \cite{zuo2022constructing}. The input of $C$ is the generated SFC and NF.

\subsection{Loss Functions}
The Kullback-Leibler (KL) divergence and reconstruct loss must be monitored during the training process to keep the VAE-based model stable and robust. The generated structural-functional connectivity must be discriminative after disentangling and fusing the fMRI and DTI. We design four hybrid loss functions: the KL loss ($L_{kl}$), the reconstruct loss ($L_{rec}$), the distangled cosine distance loss ($L_{cos}$), and the classification loss ($L_{cls}$). They are defined as follows:

\begin{equation}
\begin{split}
L_{kl} &= KL(\bm{Z_{ss}}| \mathcal {N}(0,1)) + KL(\bm{Z_{su}}| \mathcal {N}(0,1)) \\
&+ KL(\bm{Z_{ff}}| \mathcal {N}(0,1)) + KL(\bm{Z_{fu}}| \mathcal {N}(0,1))
\end{split}
\end{equation}

\begin{equation}
L_{rec} = || \bm{A_1'} - \bm{A_1} ||_2 + || \bm{A_2'} - \bm{A_2} ||_2
\end{equation}

\begin{equation}
L_{cos} = \frac{\bm{Z_{su}} \cdot \bm{Z_{fu}}}{||\bm{Z_{su}}|| \ast ||\bm{Z_{fu}}||}
\end{equation}

\begin{equation}
L_{cls} = -\bm{y} \cdot log(C(\bm{A_m}))
\end{equation}
here, $\bm{y}$ is the one-hot vector that represents the truth label.

\section{Experimental Results}

In this study, we selected subjects with both fMRI and DTI from the Alzheimer's Disease Neuroimaging Initiative (ADNI) dataset. Three stages during the AD progression are considered: normal control (NC), early mild cognitive impairment (EMCI), and late mild cognitive impairment (LMCI). To remove the impact of the imbalanced labels, 76 subjects are selected for each stage. The GRETNA and PANDA toolboxes are utilized to preprocess the fMRI and DTI, respectively. Detailed procedures are described in the work \cite{zuo2021prior}. The final outputs of the pre-processing operation are FC and SC.

The model is trained on the Ubuntu 18.04 platform with the TensorFlow tools. The optimization algorithm is Adam, where the weight decay and momentum rates are 0.01 and (0.9, 0.99). Two binary classification tasks (i.e., NC vs. EMCI and EMCI vs. LMCI) are conducted to evaluate the model's performance. Three methods are introduced to compare the classification performance of FC and SC. These methods are as follows: (1) DCNN\cite{atwood2016diffusion}, (2) MVGCN\cite{zhang2018multi}, (3) JNML\cite{LeiCheng2020}, and (4) Ours.

\begin{table}[]
	\renewcommand\arraystretch{1.}
	\setlength{\abovecaptionskip}{0pt}%
	\setlength{\belowcaptionskip}{10pt}%
	\caption{Comparison of classification performance using different fMRI-DTI fusing methods(\%).}
	\label{tab1}
    \centering
	\begin{tabular}{c|cccc|cccc}
		\hline
		\multirow{2}{*}{Methods} & \multicolumn{4}{c|}{NC vs. EMCI}                                  & \multicolumn{4}{c}{EMCI vs. LMCI}                                 \\ \cline{2-9}
		& ACC            & SEN            & SPE            & F1             & ACC            & SEN            & SPE            & F1             \\ \hline
		DCNN                     & 83.55          & 84.21          & 82.89          & 83.66          & 87.50          & 86.84          & 88.15          & 87.41          \\
		MVGCN                    & 87.50          & 88.15          & 86.84          & 87.58          & 90.78          & 89.47          & 92.10          & 90.66          \\
		JNML                     & 88.15          & 89.47          & 86.84          & 88.31          & 92.10          & 90.78          & 93.42          & 92.00          \\
		\textbf{Ours}            & \textbf{90.78} & \textbf{92.10} & \textbf{89.47} & \textbf{90.90} & \textbf{93.42} & \textbf{92.10} & \textbf{94.73} & \textbf{93.33} \\ \hline
	\end{tabular}
\end{table}

\begin{figure}[htbp]
    \centering
	\includegraphics[width=0.9\textwidth]{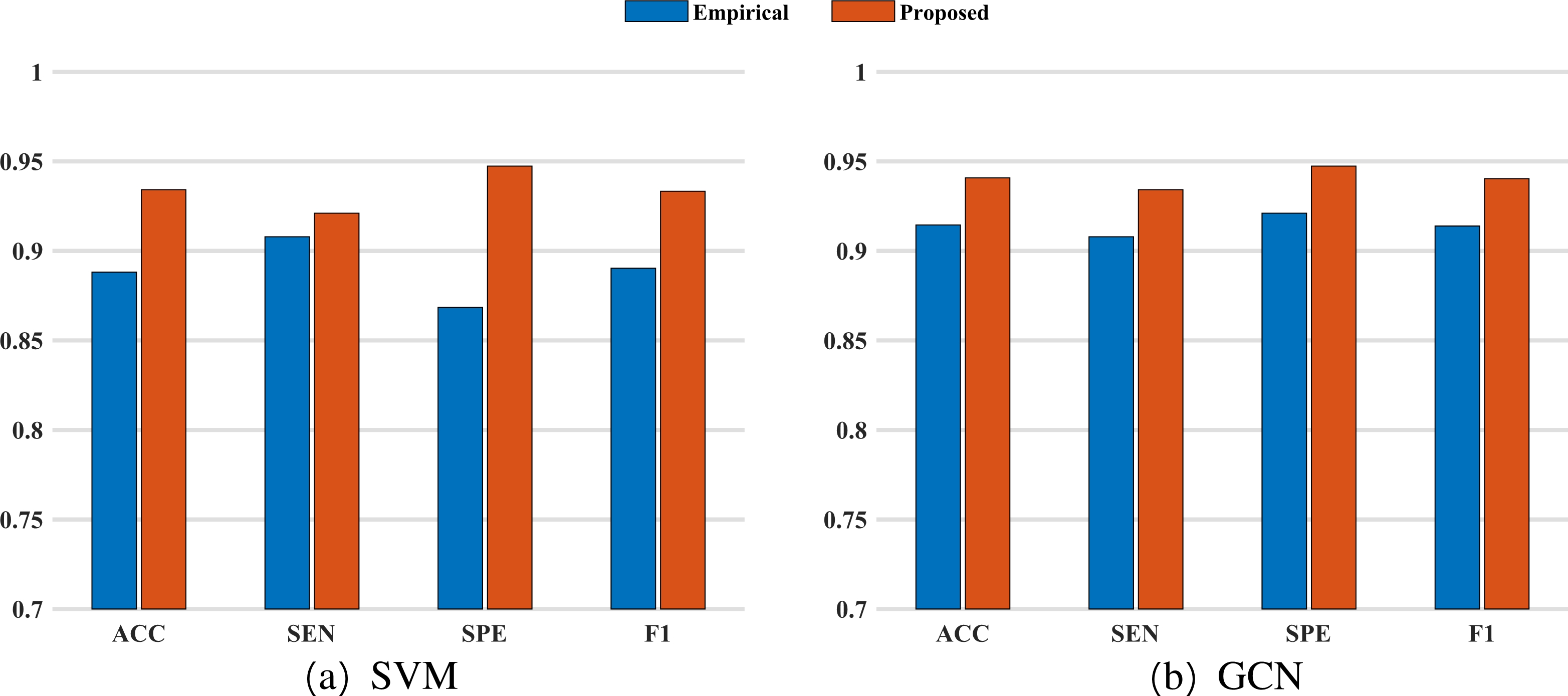}
	\caption{Classification comparison between the generated and empirical SFCs using (a) SVM classifier, and (b) GCN classifier.} \label{fig2}
\end{figure}

The classification results are presented in Table~\ref{tab1}. Our model achieves the best classification performance among the compared methods. The best results for NC vs. EMCI are an ACC value of 90.78\%, SEN value of 92.10\%, SPE value of 89.47\%, and a F1 value of 90.90\%; the task of EMCI vs. LMCI yields the best results in terms of ACC (93.42\%), SEN (92.10\%), SPE (94.73\%), and F1 (93.33\%). The same phenomenon can be observed by comparing the generated SFC and the empirical SFC. As shown in Fig.~\ref{fig2}, the generated SFCs are classified more precisely than the empirical SFCs.

\begin{figure}[htbp]
	\centering
	\includegraphics[width=0.6\textwidth]{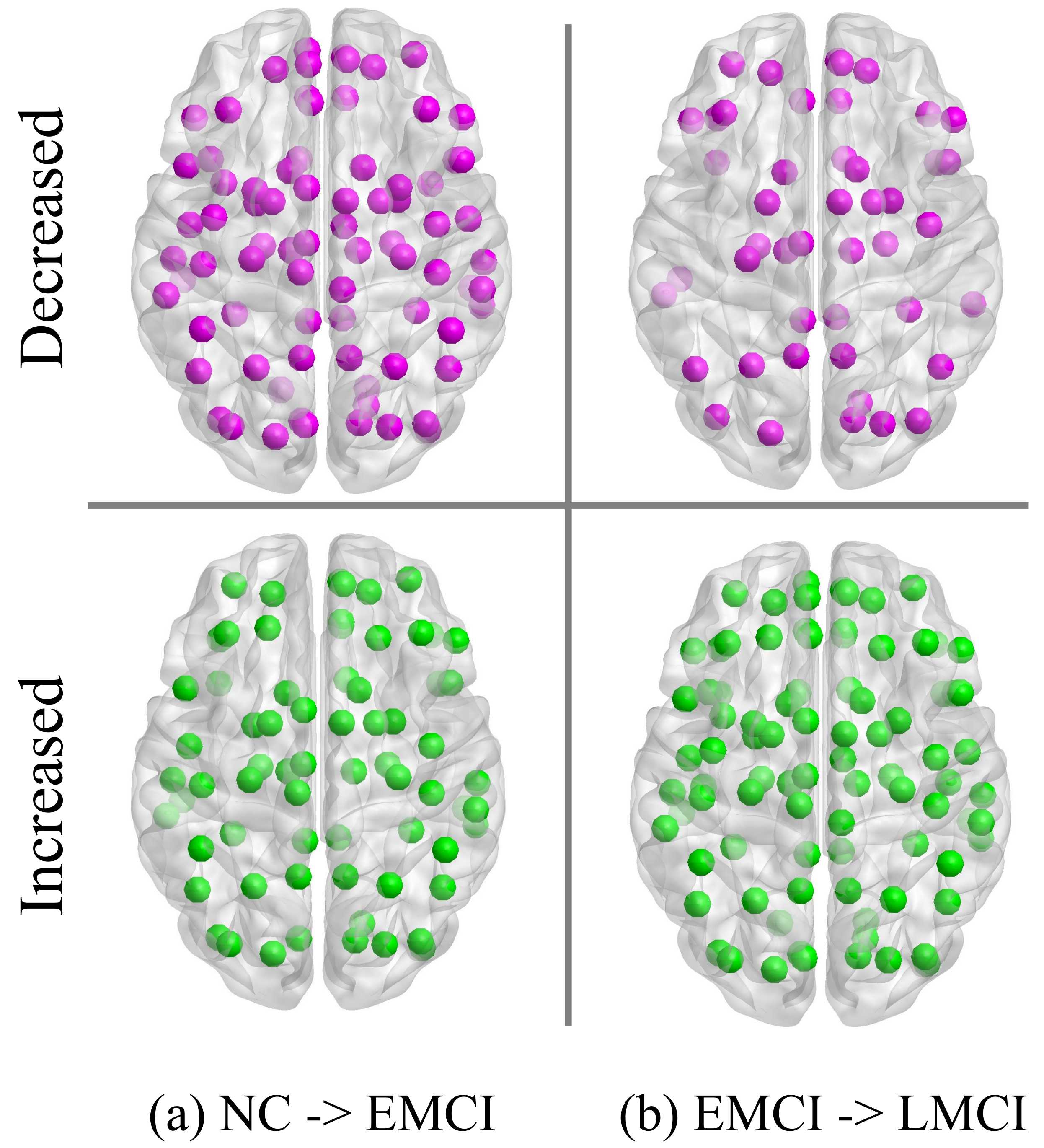}
	\caption{Spatial distribution of important connectivity-related ROIs at different stages of MCI.} \label{fig3}
\end{figure}

\begin{figure}[htbp]
	\centering
	\includegraphics[width=\textwidth]{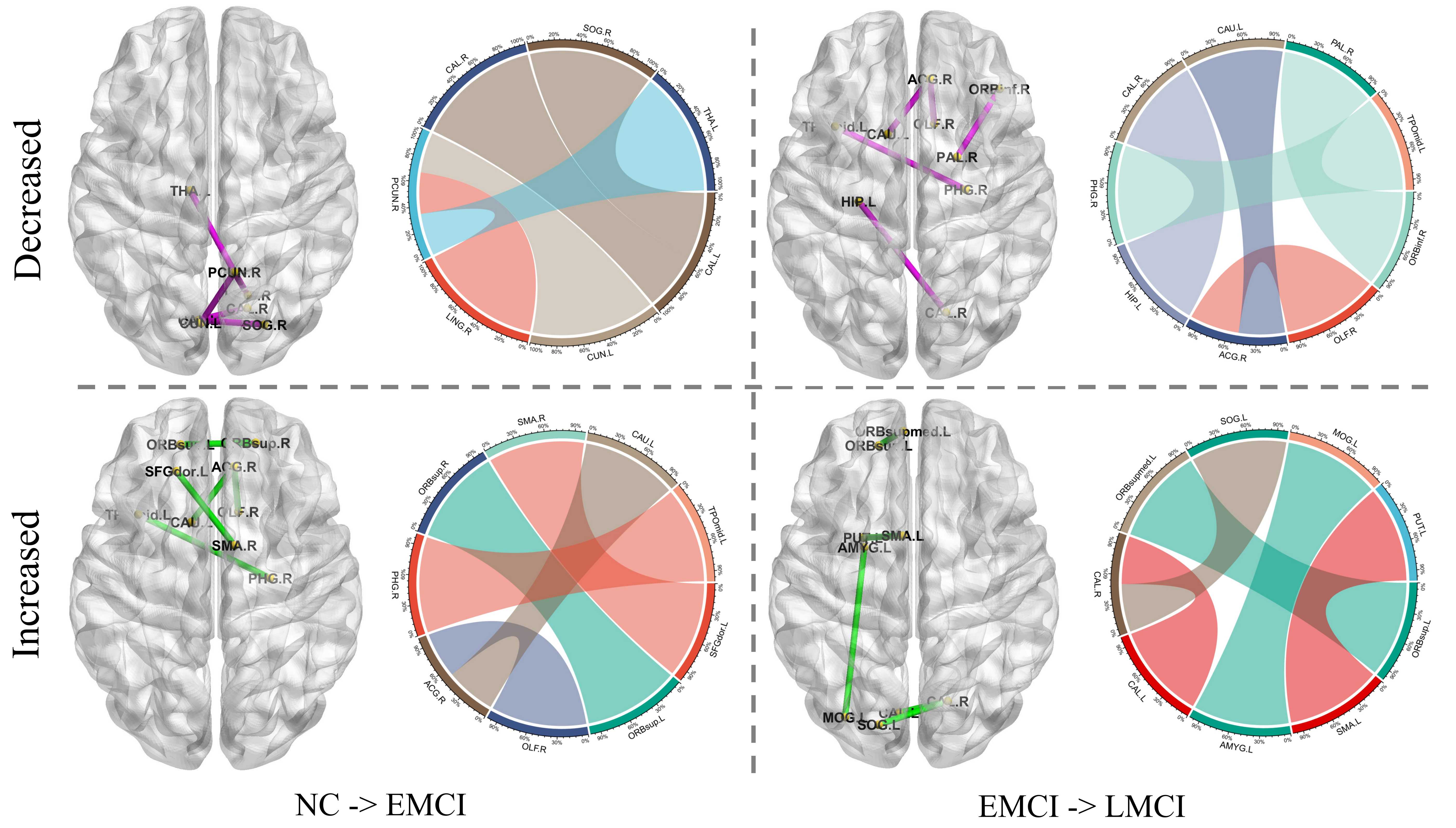}
	\caption{Qualitative and quantitative visualization of top 5 decreased and increased connections.} \label{fig4}
\end{figure}

To analyze the MCI-related brain regions and connections, we average the generated SFCs for each group (i.e., NC, EMCI, and LMCI) and compute the connectivity difference between adjacent stages. Positive values indicate increased brain connections, and negative values indicate decreased brain connections. We then select important connections by setting a threshold of 75 percent quantile. These connectivity-related ROIs are displayed in Fig.~\ref{fig3}. It shows some abnormal ROI distribution patterns when the LMCI stage occurs. In Fig.~\ref{fig4}, the top five connections in both decreased and increased situations are presented. The left of each subplot is a qualitative view, and the right of each subplot is a quantitative view with altered connection strength. From NC to EMCI, the top five increased connections are: ORBsup.L - ORBsup.R, PHG.R - TPOmid.L, ACG.R - CAU.L, SFGdor.L - SMA.R, and OLF.R - ACG.R; the top five decreased connections are: CAL.L - SOG.R, PCUN.R - THA.L, CAL.L - CAL.R, LING.R - PCUN.R, and CUN.L - PCUN.R. Patients converting from EMCI to LMCI are likely to lose the following five connections: OLF.R - ACG.R, PHG.R - TPOmid.L, ACG.R - CAU.L, HIP.L - CAL.R, and ORBinf.R - PAL.R, while five other connections may be increased: ORBsup.L - ORBsupmed.L, AMYG.L - MOG.L, SMA.L - PUT.L, CAL.R - SOG.L, and CAL.L - CAL.R.

\section{Conclusion}

This study proposes a model named Hierarchical Structural-functional Connectivity Fusing (HSCF) to build brain structural-functional connectivity matrices and forecast abnormal brain connections by inputting fMRI and DTI. In particular, the graph convolutional networks incorporate prior knowledge into the separators for disentangling each modal information. The additional HRF module can maximize the integration of pertinent and useful data across modalities, which makes the generated structural-functional connectivity more reliable and discriminative in the analysis of cognitive diseases. Study results conducted on the public ADNI database reveal the proposed model's effectiveness in classification evaluation. The identified abnormal connections will likely be biomarkers for cognitive disease study and treatment.

\subsubsection{Acknowledgements} This work was supported in part by the Guangdong Social Science Planning Project under Grant GD21YWW01, in part by the Special Fund for Young Teachers in Basic Scientific Research Business  Expenses of Central Universities under Grant 2023qntd28, in part by the Young Doctoral Research Start-Up Fund of Hubei University of Economics under Grant XJ22BS28.

%
%
%
%

\end{document}